\documentstyle[aps,floats,prl,twocolumn,epsf]{revtex}

\begin{document} 

\title{Theory of measuring the ``Luttinger-g'' of a
  one-dimensional quantum dot}  
\author{T. Kleimann$^{1}$, F. Cavaliere$^{1}$, M.
  Sassetti$^{1}$ and B. Kramer$^{2}$
  \vspace{1mm}\\
  $^{1}$ Dipartimento di Fisica, INFM, Universit\`{a} di Genova,
  Via Dodecaneso 33, I-16146 Genova\\
  $^{2}$I. Institut f\"ur Theoretische Physik, Universit\"at Hamburg,
  Jungiusstra\ss{}e 9, D-20355 Hamburg}
\author{ {\small (May 06, 2002)}
\vspace{3mm}
}
\author{
\parbox{14cm}
{\parindent4mm
\baselineskip11pt
{\small We study electron transport through a quantum dot in a
  Tomonaga-Luttinger liquid with an inhomogeneity induced either by a
  non-uniform electron interaction or by the presence of tunnel resistances of
  contacts. The non-analytic temperature behavior of the conductance peaks
  show crossovers determined by the different energy scales associated with
  the dot and the inhomogeneity despite the Coulomb blockade remains intact.
  This suggests an explanation of recent findings in semiconductor quantum
  wires and carbon nanotubes.}
\vspace{4mm}
}
}
\author{
\parbox{14cm}{
}
}
\maketitle

One dimensional (1D) electron systems are important paradigms for studying the
effects of impurities and interactions in condensed matter. Here,
electron-electron interaction can be treated by the bosonization
technique using the Tomonaga-Luttinger liquid (TLL) model. The energetically
lowest excitations are collective charge- and spin-density waves \cite{v95}.
Correlation functions can be exactly obtained. As a function of
temperature, frequency and/or bias voltage, non-analytic power laws of the
form $C(\tau)\propto \tau^{\nu(g)}$ have been predicted, where $\nu(g)$ is a
non-rational exponent that contains the interaction parameter $g$. Since TLL
appears to be of fundamental importance in modern condensed matter physics,
experimentally confirming such behavior is very important.

The search for ``Tomonaga-Luttinger behavior'' has been very intense during
recent years. In tunable semiconductor quantum wires contacted to nearly
adiabatic funnels \cite{ta95} conductance quantization was found to be weakly
affected by electron correlations. It was argued that in these systems the
conductance of the leads (assumed as Fermi liquids) is measured \cite{s96}.
Non-universal conductance quantization was detected in semiconductor
cleaved-edge-overgrowth (CEO) quantum wires \cite{ya96}. This was assigned to
electron backscattering in the contacts \cite{p00}.  Theoretical approaches
considered contacts and impurity scattering \cite{m95,f96,l01}. In recent
CEO-experiments, a 1D quantum dot was fabricated between two impurities
\cite{a00}. The dependence on temperature $T$ of Coulomb blockade peaks was
cleanly analyzed within the TLL model \cite{f98,b00,k00}.  Most strikingly,
interaction parameters deduced from different measured quantities were found
to be inconsistent: from the charging energy $E_{\rm c}$, $g$ was estimated to
be a factor of two smaller than the value obtained from the temperature
dependence of the conductance peaks. Similar inconsistencies were observed in
the temperature behavior of the conductance of a carbon nanotube (CN)
SET-transistor formed between two buckles \cite{p01}.

In this paper, we consider a 1D quantum dot in a TLL with an inhomogeneity
induced either by a non-uniform interaction parameter or by the presence of
contacts modeled by tunnel resistances. First, in view of the CEO-system, we
determine the conductance for $g(x)$ interpolating smoothly between
$g_{\infty}=g(|x|\to\infty)$ and $g_{\rm b}=g(x_{\rm b})$ in the dot. We find
that the interaction in different regions of the system can be probed. While
$E_{\rm c}$ and the level spacing $\varepsilon$ in the dot are determined by
the interaction near $x_{\rm b}$, the temperature behavior of the conductance
peaks show crossovers between $g_{\infty}$ (low $T$) and $g_{\rm b}$ (high
$T$).  Second, we determine the conductance of a quantum dot embedded in a TLL
with $g={\rm const}$ but attached to metallic contacts. This is appropriate
for CNs. We show that the temperature behavior of the Coulomb peaks can be
entirely determined by the non-analytic powers laws induced by the contacts
though the Coulomb blockade peaks remain perfectly intact. Quite generally,
the crossovers can be traced back to the competing energy scales associated
with the quantum dot and the inhomogeneity. Our results not only put earlier
theoretical findings into a general frame but also explain quantitatively the
--- at the first sight --- discrepant findings in the recent experiments which
seems to us of outstanding importance after the intense search for
TLL-behavior during more than three decades.

We consider the Hamiltonian $H=H_{0}+H_{\rm b}$ with
\begin{equation}
  \label{eq:1}
  H_{0}=\frac{\hbar v_{\rm F}}{2}\int {\rm d}x\,\left\{\Pi^{2}(x)
+\frac{1}{g^{2}(x)}[\partial_{x}\vartheta(x)]^{2}\right\}
\end{equation}
describing the clean TLL-system with the conjugate Boson
fields $[\Pi(x),\vartheta(x')]=i\delta(x-x')$ and a local
inhomogeneous density-density interaction
\begin{equation}
  \label{eq:2}
\frac{1}{g^{2}(x)}=
\frac{1}{g_{\infty}^{2}}+\frac{\varphi(x)}{\pi\hbar v_{\rm F}}.  
\end{equation}
We assume $\varphi$ as smooth, with a maximum of width $L^{*}\gg
a$ near the impurities, and $\varphi(x)\to 0$ when $|x|\to\infty$.

The contribution of the impurities located at $x_{\rm b}^{\pm}=x_{\rm b}\pm
a/2$ is $ H_{\rm b}= U_{\rm b}\cos(\pi N_{+}) \cos[\pi(n_0+N_{-})]$ where
$N_{\pm}=[\vartheta(x_{\rm b}^{+})\pm\vartheta(x_{\rm b}^{-})]/\sqrt{\pi}$.
The quantity $N_{+}$ and $N_{-}$ are associated with the unbalanced particles
between left and right leads, and the fluctuations of the particle number in
the dot with respect to the mean value $n_0=k_{\rm F}a/\pi$, respectively
($k_{\rm F}$ Fermi wave number). The charge modes of the inhomogeneous TLL act
on $N_{\pm}$ as an external Bosonic bath. Their influence can be exactly
evaluated in terms of a dissipative impedance \cite{k00}. For $L^{*}>a$, such
that near the quantum dot $g(x)\approx g_{\rm b}$, the resulting Euclidean
effective action is
\begin{equation}
S_{\rm eff}[N_{+},N_{-}]=\frac{k_{\rm B}T}{2}\sum_{r=\pm,n}
N_{r}(\omega_n)K_{r}(\omega_n)N_{r}(\omega_n).
\end{equation} 
The Fourier transforms of the dissipative kernels $K_{r}$, at Matsubara
frequencies $\hbar\omega_n=2k_{\rm B}T\pi n$, are
\begin{equation}
  \label{eq:9}
K_{\pm}(\omega_{n})=\frac{\pi}{2}\left[
{\cal G}(\omega_{n};x_{\rm b},x_{\rm b})\pm 
{\cal G}(\omega_{n};x_{\rm b}^{+},x_{\rm b}^{-})\right]^{-1}
\,.
\end{equation}
The zero-frequency limit of $K_{-}$ gives the charging energy, $E_{\rm
  c}=\hbar K_{-}(\omega_{n}\to 0)/2$. The dissipative effects of the
collective modes are described by the spectral density
\begin{equation}
  \label{spectral}
  J_{\rm tot}(\omega)=\frac{1}{\pi}
{\rm Im}[K_{-}(\omega_{n}\to +i\omega)+
K_{+}(\omega_{n}\to +i\omega)].
\end{equation}

These quantities depend on the time ordered propagator ${\cal
  G}(\tau;x,x')\equiv \langle{\bf
  T_{\tau}}\vartheta(x,\tau)\vartheta(x',0)\rangle$
defined with respect to $H_0$ via the resolvent equation
\begin{equation}
  \label{eq:3}
  \left[\frac{\omega^{2}_{n}}{v_{\rm F}}-\frac{\partial}{\partial x}
\frac{v_{\rm F}}{g^{2}(x)}\frac{\partial}{\partial x}
\right]{\cal G}(\omega_{n};x,x')=\delta(x-x').
\end{equation}
We solve (\ref{eq:3}) linearly in $\varphi(x)$
($x>x'$, $\omega_{n}>0$),
\begin{eqnarray}
  \label{gsolution}
{\cal G}(\omega_{n};x,x')&=&\frac{g_{\infty}}{2\omega_{n}}
\Big\{e^{-\eta |x-x'|}-C\Big[e^{-\eta(x+x')}\times\Big.\nonumber\\
&&\hspace{-1.5cm}\times\int_{-\infty}^{x'}{\rm
      d}y\,\varphi(y)e^{2\eta y}\Big.
+e^{\eta(x+x')}\int_{x}^{\infty}{\rm
      d}y\,\varphi(y)e^{-2\eta y}\nonumber\\
&&\Big.\Big.\qquad\qquad-e^{-\eta(x-x')}\int_{x'}^{x}{\rm
      d}y\,\varphi(y)\Big]\Big\}
\end{eqnarray}
with the constants $\eta=g_{\infty}\omega_{n}/v_{\rm F}$
and $C=g_{\infty}^{3}\omega_{n}/2\pi\hbar v^{2}_{\rm F}$.

\begin{figure}[htbp]
\setlength{\unitlength}{1cm}
\begin{picture}(7.0,3)(0,-0.4)
\put(0.0,-1.7){\epsfxsize=8.5cm
             \epsffile{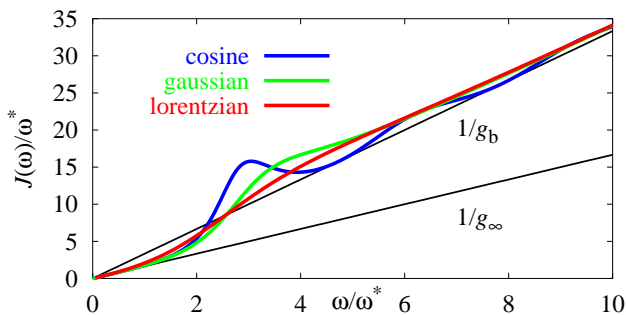}}
\end{picture}\par
\vskip1.7cm
\caption{The spectral density
  $J(\omega)/\omega^{*}$ for different interactions centered around $x_{\rm
    b}=0$ (see text). Black lines: asymptotic behaviors with slopes given by
  $g_{\infty}=0.6$ and $g_{\rm b}=0.3$.}
  \label{fig:1}
\end{figure}

Without impurities, one obtains from the above Green function the linear
dc-conductance $G_{0}=2e^{2}\omega_{n}{\cal G}(\omega_{n}\to
0)/h=g_{\infty}e^{2}/h$, consistent with earlier work \cite{s96}, and
indicating that the linear conductance is a {\em global probe}. For the
quantum dot, the results are more complex. On the one hand, by expanding
(\ref{eq:9}) and (\ref{gsolution}) consistently to the order $a/L^{*}\ll 1$,
the charging energy turns out to be $E_{\rm c}=\pi\hbar v_{\rm F}/2ag^{2}_{\rm
  b}$. It is determined by the interaction at the position of the quantum dot
and is a {\em local probe}.  On the other hand, the spectral density
(\ref{spectral}) can be decomposed in two parts that describe the influences
of the leads, and of the energy discretization in the dot, $J_{\rm
  tot}(\omega)=J_{\rm l}(\omega)+J_{\rm d}(\omega)=J(\omega)
[1+\varepsilon\sum_{n=1}^{\infty}\delta(\hbar\omega-n\varepsilon)]$, where
$J(\omega)={\rm Im}[{\cal G}^{-1}(i\omega;x_{\rm b},x_{\rm b})]/2$ and
$\varepsilon\equiv 2g_{\rm b}E_{\rm c}$ the discretization energy of the
plasmon modes in the quantum dot. The function $J(\omega)$ is shown in
Fig.~\ref{fig:1} for $\varphi(x)=\exp{[-(2x/L^{*})^2]}$;
$[1+(2x/L^{*})^2]^{-1}$; and $[1+\cos(\pi x/L^{*})]/2$, $|x|<L^{*}$. There is
a crossover between low- and high-frequency behavior, independently of the
particular interaction, namely $J(\omega\ll\omega^{*})=\omega/g_{\infty}$ and
$J(\omega\gg\omega^{*})=\omega/g_{\rm b}$, where $\omega^{*}=v_{\rm
  F}/g_{\infty}L^{*}$ is the crossover frequency corresponding to the
characteristic length scale of the inhomogeneity.

The conductance $G_{\rm d}$ in the region of the Coulomb blockade is
calculated by using sequential tunneling through high barriers in the presence
of an external gate voltage $V_{\rm g}$ \cite{f98,b00}. The latter defines the
reference particle number in the dot. We consider $k_{\rm B}T\ll
\varepsilon$. The dependence on the gate energy $\mu=e(V_{\rm g}-V^{\rm
  res}_{\rm g})$, that measures the shift with respect to the resonance value
$eV_{\rm g}^{\rm res}=E_{\rm c}[2(n-n_0)+1]$, and the temperature is
\begin{equation}
  \label{eq:11b}
G_{\rm d}=
\frac{e^{2}\,e^{-\mu/2k_{\rm B}T}}
{4k_{\rm B}T\cosh(\mu/2k_{\rm B}T)}\,
w_0(\varepsilon,T)\,\gamma(\mu,T).
\end{equation}
Here $\gamma(\mu,T)=(\Delta^{2}/4)\int {\rm d}t\,\exp{[i\mu t/\hbar-W_{\rm
    l}(t)}]$ is a tunneling rate through a single impurity with tunneling
frequency $\Delta$. The temperature and time dependent dissipative kernel
$W_{\rm l}(t)$ is given as an integral that contains the above spectral
density $J(\omega)$, with a frequency cutoff $\omega_{\rm c}$. The weight
$w_0(\varepsilon,T)=(\varepsilon/h)\int_{0}^{h/\varepsilon} {\rm
  d}t\,\exp{[-W_{\rm d}(t)}]$ is the time average of the periodic kernel of
the dot $W_{\rm d}(t)$ \cite{k00}. Higher components that exhibit the
excitation spectra do not contribute in the linear regime since $k_{\rm
  B}T<\varepsilon$. Below we present numerical results obtained for the
Lorentzian inhomogeneity (Fig. \ref{fig:1}).

\begin{figure}[t]
\vskip2.7cm
\setlength{\unitlength}{1.1cm}
\begin{picture}(7.0,3.5)(0,-0.4)
\put(-0.0,1.5)
{\epsfxsize=9.5cm\epsfysize=7.5cm
              \epsffile{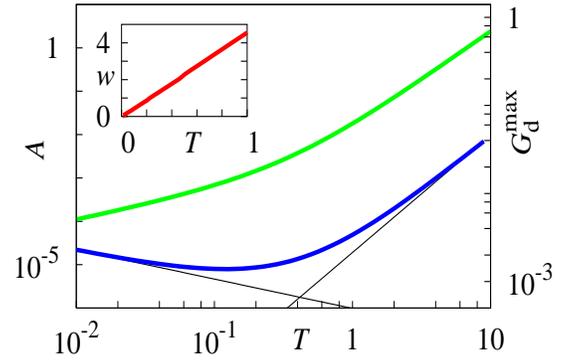}}
\end{picture}
\setlength{\unitlength}{1.0cm}
\begin{picture}(7.,5.5)(0,-0.4)
\put(1.3,1.2)
{\epsfxsize=7cm\epsfysize=5cm
 \epsffile{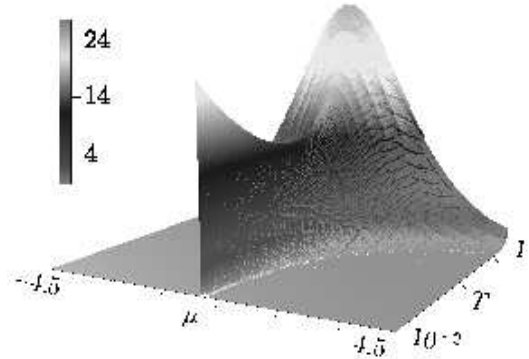}}
\end{picture}
\vskip-1cm 
\caption{Conductance peak of a 1D quantum dot for 
  lorentzian interaction ($x_{\rm b}=0$, $g_{\infty}=0.6$, $g_{\rm b}=0.3$,
  $\omega^{*}/\omega_{\rm c}=10^{-3}$). Top: double logarithmic plot as a
  function of the temperature $T$ in units of $T^{*}=\mu^{*}/k_{\rm B}$
  ($\mu^{*}=\hbar\omega^{*}$) of the peak height $G^{\rm max}_{\rm d}$ (blue
  curve, right scale, units $G_{0}=(\Delta/4\omega_{\rm
    c})^{2}(\varepsilon/\hbar\omega_{\rm c})^{1/g_{\rm b}}e^{2}/\hbar$), and
  the peak area $A$ normalized to $G_{0}\mu^{*}$ (green curve, left scale);
  inset: peak width $W$ (units $\mu^{*}$) as a function of $T$. Bottom:
  conductance as a function of the gate energy $\mu$ and of the temperature
  $T$ (units $\mu^{*}$, $T^{*}$), color code (left, units $10^{-4}G_{0}$).}
  \label{fig:2}
\end{figure}

The conductance for a single Coulomb blockade peak as a function of
temperature and gate energy is shown in Fig. \ref{fig:2} for $g_{\infty}=0.6$
and $g_{\rm b}=0.3$.  Independently of the inhomogeneity, the width $w$ of the
peak increases linearly with temperature. This implies that even in the
inhomogeneous case the area $A$ under the peak and the peak height are
connected by $A\propto T G^{\rm max}_{\rm d}$. The maximum of the peak has a
{\em non-monotonous} behavior in temperature with a minimum around the
crossover that corresponds to the saddle in the 3D plot. The temperature
dependence of the maximum can be written as $ G^{\rm max}_{\rm d}(T)\propto
T^{1/\tilde{g}(T)-2} $ where $\tilde{g}(T)$ shows a crossover between
$\tilde{g}(T\ll T^{*})=g_{\infty}$ and $\tilde{g}(T\gg T^{*})=g_{\rm b}$ with
$k_{\rm B}T^{*}=\hbar\omega^{*}$. The non-monotonous behavior of the
conductance is due to the particular choice of the interaction parameters
$g_{\rm b}<1/2<g_{\infty}$. If both are larger or smaller than the critical
value $1/2$ the conductance decreases or increases monotonously with
temperature, respectively. However, independently of the details near the
crossover, the high- and low-temperature behaviors of the peaks are always
dominated by the local and global properties of the interaction represented by
$g_{\rm b}$ and $g_{\infty}$, respectively. This means that a measurement of
the linear conductance of the 1D quantum dot at low temperatures reflects the
interaction far away from the barrier, and therefore is a {\em global probe}.
On the other hand, when measuring the Coulomb peak at higher temperatures,
$T^{*}<T$, the interaction close to the dot will dominate. In this region, the
experiment is a {\em local probe} for the interaction.

\begin{figure}[t]
  \setlength{\unitlength}{1cm}
\begin{picture}(7.0,5.0)(0,-0.4)
\put(-0.1,-0.5)
{\epsfxsize=7.5cm\epsfysize=6.0cm
              \epsffile{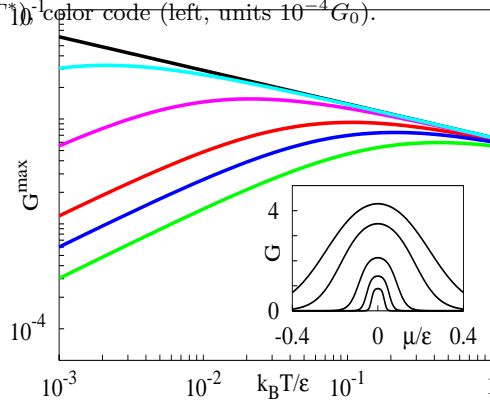}}
\end{picture}
\vskip0.2cm\par
  \caption{Double logarithmic plot of the  height  of the conductance peak
    of a quantum dot (units $e^2/h$) connected to leads via resistive tunnel
    contacts as a function of temperature $k_{\rm B}T/\varepsilon$ with
    $2R_{\rm c}/R_{\rm d}= 0, 10^{-3}, 10^{-2},5\cdot 10^{-2}, 10^{-1},2\cdot
    10^{-1}$ (top to bottom), $\alpha\equiv -1+1/g=0.68$ and $R_{\rm d}=150
    h/e^2$. Inset: conductance $G$ in units of $10^{-3}e^{2}/h$ for $R_{\rm
      c}/R_{\rm d}=0.1$ (green curve of main figure) as a function of
    $\mu/\varepsilon$; temperatures $10^{2}T/\varepsilon=8,5,2,1,0.5$ (top to
    bottom).}
  \label{fig:3}
\end{figure}

That contacts may become crucial when interpreting experimental data can also
be seen by considering a TLL model with homogeneous interaction, $g(x)=g$, but
its ends connected to a normal metallic lead via point like tunnel contacts.
The temperature dependence of the contact conductance is \cite{k92}
\begin{equation}
  \label{contact}
G_{\rm c}(T)=\frac{1}{R_{\rm c}}\,
\left(\frac{k_{\rm B}T}{\varepsilon}\right)^{\alpha}\,, \qquad\alpha =
g^{-1}-1
\end{equation}
with the prefactor $R_{\rm c}^{-1}$ containing the tunneling resistance
between the lead and the TLL. For later comparison with the dot conductance we
have chosen the discretization energy $\varepsilon$ as the energy scale,
including in $R_{\rm c}$ the rescaling between $\omega_{\rm c}$ and
$\varepsilon$. If a quantum dot is embedded into the TLL with complete
momentum randomization in the contacts, the total resistance can be obtained
by adding the resistances of the contacts (\ref{contact}), and that of the
quantum dot (\ref{eq:11b}) with $g={\rm const}$: $G^{-1}(T)\equiv R(T)=G_{\rm
  c}^{-1}(T)+G_{\rm d}^{-1}(T)$. This can still yield Coulomb blockade
oscillations (Fig.~\ref{fig:3}) but with a crossover in the temperature
behavior. Near the maximum of a Coulomb peak the dot conductance
scales as
\begin{equation}
  \label{eq:11}
G^{\rm max}_{\rm d}(T)=\frac{1}{R_{\rm
  d}}\left(\frac{k_{\rm B}T}{\varepsilon}\right)^{\alpha-1}\,.
\end{equation}
In this region, the $G(T)$ will be determined by $G_{\rm c}(T)$ for
temperatures lower than a crossover value that depends on the ratio $R_{\rm
  c}/R_{\rm d}$. On the other hand, in the tails of the peak, the $G_{\rm
  d}(T)$ always dominates, since there in any case $G_{\rm d}(T)\ll G_{\rm
  c}(T)$. The peak height $G^{\rm max}(T)$ shows a crossover between global
(small $T$) and local (large $T$) power laws. For $\alpha<1$, it has a maximum
near the crossover temperature is $T^{*}=2R_{\rm c}\varepsilon/R_{\rm d}k_{\rm
  B}$. For $R_{\rm c}=0$ (no contact) only the power law behavior of the dot
is obtained (black curve in Fig.~\ref{fig:3}).

The above results suggest that inhomogeneities cannot be neglected when
deducing the interaction parameter in experimental data, especially in the
presence of backscattering impurities. In \cite{a00} the charging energy of a
1D dot embedded in a CEO-quantum wire has been determined from the distance
between Coulomb peaks, $E_{\rm c}=2.2\,$meV with an estimate of $a\approx
100\ldots 200\,$nm. It has been found that $g_{\rm b}\approx 0.4$ \cite{k00}.
Measuring the temperature dependence of two conductance peaks in the range
$T\approx 0.2\ldots 2\,$K, a value $g_{\rm exp}\approx 0.8$ has been estimated
(without spin) \cite{a00}. Taking into account the spin reduces the latter
value slightly to $\approx 0.7$ but does not solve the inconsistency. Given
the good fit of the experimental data for $T<2\,$K to the power law with a
single value for $g_{\rm exp}>g_{\rm b}$ we conclude $g_{\rm exp}\approx
g_{\infty}$, and that $T^{*}$ should be larger than $2\,$K.  By inspection of
Fig. \ref{fig:2} we estimate $T^{*}\approx 10\,$K. This value certainly
depends on the shape of the inhomogeneity, but should be of the correct order.
With $v_{\rm F}\approx 10^{5}\,$m/s we find $L^{*}\approx 100\,$nm. This is
not larger than the estimated length of the 1D quantum dot.  But in view of
the above idealized model assumptions, we can conclude that the temperature
dependence of the conductance peaks are described by $g_{\infty}$, and not by
$g_{\rm b}$.  Our result could be further experimentally addressed by changing
the parameters of the CEO-quantum wires, which should influence especially
$L^{*}$.

The influence of the contacts must also be present in the temperature behavior
of the Coulomb blockade peaks of a quantum dot created by two buckles in a CN.
In the absence of the dot a CN with ends attached to Au-wires shows the
contact conductance $G_{\rm c}(T)$ of (\ref{contact}) with $\alpha=(g_{\rm
  nano}^{-1}-1)/4$ and $g_{\rm nano}\approx 0.28$ \cite{b99}. The different
relation between $\alpha$ and $g_{\rm nano}$ from the one given in
(\ref{contact}) is due to the presence of further three non-interacting modes.
In the presence of the dot, with the experimental value $\alpha=0.68$
(corresponding to $g_{\rm nano}=0.27$ \cite{p01}) we obtain from (\ref{eq:11})
that $G_{\rm d}^{\rm max}(T)$ increases with decreasing $T$ (black curve in
Fig.~\ref{fig:3}). On the other hand, the maximum $G^{\rm max}$ of the {\em
  total} conductance including the contacts decreases with decreasing $T$
according to the power law of the contact conductance for $T<T^{*}$ and for a
suitable ratio $R_{\rm c}/R_{\rm d}$.  This is consistent with the
experimental data and explains the discrepant behavior between the observed
$G_{\rm max}(T)$ and the dot conductance in the sequential tunneling regime
\cite{p01}.

In conclusion, we have investigated transport in a 1D quantum dot embedded in
a TLL with an inhomogeneity either due to non-uniform interaction or to the
presence of contacts. We have identified quantities that measure the
interaction locally and globally. We predict a crossover in the temperature
behavior of Coulomb blockade peaks between regions that probe the interaction
far away from and close to the dot for $T\ll T^{*}$, and $T\gg T^{*}$,
respectively, though the Coulomb blockade remains intact. For the
non-homogeneous interaction, the crossover is determined by the characteristic
length of the region where $g(x)\approx {\rm const} \equiv g(x_{\rm b})$. When
assuming a homogeneous interaction strength, the crossover is determined by
the ratio between contact and dot resistances. Quite generally, the behavior
can be understood by considering the energy scales set by the quantum dot and
the inhomogeneity: if $E_{\rm c}\gg k_{\rm B}T^{*}$ one may fulfill the
condition $k_{\rm B}T\ll E_{\rm c}$ necessary for getting Coulomb blockade
and nevertheless observe a crossover from global to local behavior in the
non-analytic power law dependence of the conductance peaks which is governed
by the energy scale of the inhomogeneity. The results are used to
understand consistently and quantitatively recent, fundamentally important
experiments done on semiconductor CEO-quantum wires and CNs.

We acknowledge discussions with M. Grifoni, F. Napoli and J.
Stockburger, and support by the EU within TMR (FMRX-CT98-0180), RTN
(HPRN-CT2000-00144), italian MURST via PRIN00, and the DFG via SFB 508
``Quan\-ten\-materialien'' of the Universit\"at Hamburg.

\end{document}